\input{psfig.sty}
\documentstyle[]{laa}    
\def\deg{$^{\circ}$}
\begin{document}
\thesaurus{20 (11.01.2 ; 11.02.1 ; 13.18.1) }
\title {Extended radio emission in BL Lac objects\\
I: the images}
\author{P. Cassaro\inst{1,5} \and C. Stanghellini\inst{1} \and M. Bondi\inst{2}
\and D. Dallacasa\inst{3} \and R. Della Ceca \inst{4} \and R.A. Zappal\`a 
\inst{5}}
\institute{
Istituto di Radioastronomia del CNR, C.P. 141, I-96017 Noto SR, Italy 
\and 
Istituto di Radioastronomia del CNR, Via Gobetti 101, I-40129 Bologna, Italy 
\and
Dipartimento di Astronomia, Universit\`a di Bologna, Via Ranzani 1, I-40127,
Bologna, Italy
\and
Osservatorio Astronomico di Brera, Via Brera 28, I-20121 Milano, Italy
\and  
Istituto di Astronomia dell'Universit\`a, Citt\`a Universitaria, viale Andrea Doria 6, I-95125 Catania, Italy}

\offprints{P.Cassaro(e-mail:cassaro@ira.noto.cnr.it)}

\date{Received 24 March 1999 / Accepted 28 July 1999}
\maketitle

\begin{abstract}

We have observed 28 sources selected from  the 1Jy sample of BL Lac objects 
(Stickel et al. 1991) with the Very Large Array (VLA) in A, B and D 
configurations at 1.36, 1.66 and 4.85 GHz, and/or with the Westerbork 
Synthesis Radio Telescope (WSRT) at 1.40 GHz. In this paper 
we present high sensitivity images at arcsecond resolution of the 18 
objects showing  extended structure in our images, and of another source from the FIRST 
(Faint Images of the Radio Sky at Twenty-cm)
survey (Becker et al. 1995).  
In general our high sensitivity images reveal an  amount of extended 
emission larger than previously reported. 
In some objects the luminosity of the extended structure 
is comparable with that of 
FR~II radio sources. 
A future paper will be devoted to the interpretation of these results.  
 
\end{abstract}

\section{Introduction}

A great effort of the recent research on Active Galactic Nuclei (AGN)
has been directed  to the development of ``Unified Schemes'': a framework wherein the observational properties of different classes of AGN can be explained as intrinsically similar objects 
seen at different orientation angles to the line of sight (see Urry \& Padovani 1995 and references
therein).

In this context, it is now widely accepted that the observed properties
of BL Lac objects are  largely due to a relativistic jet pointing in the
direction of the observer (the Beaming Model; Blandford \& Rees 1978),
implying the existence of a class of radio sources (hereafter the
parent population) intrinsically identical to BL Lacs, but with
the jets oriented at large angles with respect to the line of sight.

The nature of the parent population of the BL Lacs has been the
subject of several investigations in the past. While some authors
confirm that the majority of them are low luminosity
edge darkened FR~I radio galaxies seen along their radio axis
(see Urry \& Padovani 1995 and references
therein), others suggest that the diffuse radio emission detected
around high redshift BL Lacs is more consistent with FR~II rather than FR~I
radio galaxies (see e.g. Kollgaard et al. 1992; Murphy et al.1993).

A straightforward outcome of the beaming model is that all the
properties not depending on orientation should be shared by the BL Lac
objects and their parent population.

To address this problem we planned to compare the extended radio
luminosity of the radio selected 1 Jy BL Lac sample (Stickel et al.
1991) with that of FR~I and FR~II radio galaxies.  In fact, while the
morphology is distorted by projection effects, and a morphological
classification could be uncertain, the luminosity of the unbeamed
emission is suitable for a direct comparison with the extended emission
found in the candidate parent population.

Of the 34 BL Lac objects  belonging to the 1 Jy sample 
we have selected the 28 objects without deep radio images at arcsecond scale resolution in  the literature, and/or at the arcminute scale.
Stickel et al. (1994) added three sources to their original sample:
0218+357, 2029+121 and 2150+173. In particular 0218+357 is a well known
gravitational lens (O'Dea et al. 1992; Patnaik et al. 1993, 1995). Since our
work started when these three additional sources were not
included, and furthermore, in the
literature the '1 Jy sample of BL Lac sources' refers to the 34 objects in
Stickel et al. (1991), we will not consider the forementioned  additional
sources.

Here we present data for 28 objects observed with the VLA 
and/or the WSRT; 
preliminary results on extended emission and polarization properties for a 
small fraction of these objects were presented in Stanghellini et al. (1997).
In this paper we show the images and briefly discuss the results obtained from our observations. In a forthcoming paper we will give a full discussion of the results. 
Throughout this paper we use H$_0$ = 50 km s$^{-1}$ Mpc$^{-1}$  and q$_0$ = 0.5.

\section{Observation and data reduction}

VLA observations were carried out in 1994 and 1995 at 1.36, 1.66, 4.8 GHz in
A, B, and D configurations.
WSRT observations took place in Feb. 1994 at 1.4 GHz.
In Table 1 we report the journal of these observations.

\begin{table}
\caption{ Journal of the observations: col. [1] date of the observations, col. [2] array (A,B,D, for the VLA configurations, W for WSRT), col. [3] observing frequency, col. [4] bandwidth.}
\begin{center}
\begin{tabular}{cccc}
\hline
\hline
\noalign{\smallskip}
date & array & $\nu$(GHz) & $\Delta\nu$(MHz)\\
\noalign{\smallskip}
\hline
\noalign{\smallskip}
18/04/94 & A & 1.36 & 50\\
	& A & 1.66 & 25\\
29/04/94 & A & 1.36 & 50\\
	& A & 1.66 & 25\\
04/07/94 & B & 1.36 & 50 \\
	& B & 1.66 & 25\\
	& B & 4.88 & 50$\times$2 IFs\\
22/01/95 & D & 1.36 & 50 \\
	& D & 1.66 & 25\\
21/02/95 & D & 1.36 & 50\\
	& D & 1.66 & 25\\
02/94 & W & 1.40 & 40\\
\noalign{\smallskip}
\hline
\noalign{\smallskip}
\end{tabular}
\end{center}
\end{table}

\subsection {The VLA data}

The VLA data reduction has been performed using the Astronomical Image
Processing System (AIPS) developed at the National Radio
Astronomy Observatory (NRAO). After a standard calibration we performed 
several 
iterations of imaging and phase self-calibration and one final iteration of 
phase and amplitude self-calibration. 

We observed with the VLA in A and B configuration in order to have good
sensitivity to radio emitting regions with angular sizes up to 120
arcseconds; the high resolution provided by the A-array data allowed a
good measure of the core emission and therefore we managed to have an
accurate determination of the extended emission flux density. We also
planned to combine the data at 1.36 and 1.66 GHz to improve the uv-coverage
and increase the sensitivity, but the latter frequency was generally
affected by Radio Frequency Interferences (RFI) and we could not pursue our goal. Short B array observations
were carried out at 5 GHz to study and compare the arcsecond scale structure
detected in the L band. Finally, a few sources were observed in the D array to
search for arcminute scale emission (like in 1807+698).

At 1.36 GHz the images were obtained with a multi-field clean to remove the 
contribution of strong sources in the field. The r.m.s. noise in the final images is typically in the range 0.07$-$0.18 mJy/beam, 
close to the expected thermal noise, with the exception
of the sources with declination lower than $-20^\circ$.
The dynamic range (peak/noise) is 
between 1000$-$9000, with typical values around 5000. The images at 1.66 GHz have been 
only used to verify the results obtained at 1.36 GHz.

 We found it useful to combine L band data taken with different 
configurations only for a few sources in order to increase the sensitivity to the extended 
emission while maintaining a good resolution. 
Given that the flux density of the unresolved core may vary significantly
between the epochs of the observations in the A and B configurations, we had
to normalize the core flux density before combining the different data sets
of the same source.

From the A and B array observations we obtained two subsets, each
containing only the visibilities with common uv-coverage. Then we produced
an image  for each of these subsets and determined the peak flux densities.
We subtracted from the original data set with the strongest core a pointlike 
component with intensity corresponding to the difference between the peak 
fluxes. Finally we combined the two complete data sets. 
Only for 1807+698 we also added the data in the D configuration, using the same 
technique.

\subsection{The WSRT data}

The WSRT observations were carried out to search for arcminute scale
extended emission (like for VLA D-Array data). 

The total observing time for the  observations at the WSRT was
24 hours. Each source was observed for typically five 15-min snapshots,
inclusive of slewing times, in a range of HA in order to obtain an acceptable 
uv-coverage. The typical resolution was about 13''$\times$13''cos $\delta$ and
the r.m.s. noise in the image plane was between 0.15 and 0.3 mJy/beam. 
The data reduction has been done using the NEWSTAR package 
that allows redundancy self-calibration. The flux density scale has
been referred to Baars et al. (1977) by means of the observations of 3C286,
whose flux density was assumed to be 14.77 Jy at 1.40 GHz. 

The NEWSTAR package allows
for the removal of unresolved components (i.e. components with size much 
smaller than the observing beam) in the imaging process. This has been used 
to remove the nucleus of each BL Lac object in order to
estimate the total flux density of the extended emission.

\subsection{Additional images}

We also searched for images of the 1 Jy BL Lac sources in the NRAO VLA Sky Survey
(NVSS, Condon et al. 1998)
and in the FIRST (Becker et al. 1995) survey. Most sources
appeared pointlike, except 1514$-$241 on the  
NVSS and 0828+493 and 1418+546 on
the FIRST. These three images are presented here along with the images
obtained from our data, since they show more details than those available in
the literature.

\begin{table*}
\caption{ Image parameters. We do not present images for the 
sources marked with an asterisk because they are unresolved or because the
extended structure in our images does not improve the information
in the literature. Col. [1] IAU name, col. [2] array (A,B and D, are for the VLA configurations, W is for the WSRT). `a' and `b' means that the image is from the FIRST or NVSS,
respectively, col. [3] frequency of the observations, col. [4] beam major and minor axes, col. [5] position angle of the restoring beam, col. [6] r.m.s. noise on the image plane, col. [7] peak flux density on the image. When the core has
been subtracted, the flux density not restored in  the image is shown in
parentheses.}
\begin{center}
\begin{tabular}{lcccrrl}
~~~~~~[1]&[2]&[3]&[4]&[5]&[6]~~&[7]~~\\
\noalign{\smallskip}
\hline
\hline
\noalign{\smallskip}
~~~~name & array & $\nu$  & beam & PA & r.m.s. & peak \\
& & (GHz) & (arcsec) & ($^{o}$) & (mJy/beam) & (mJy/beam)\\
\noalign{\smallskip}
\hline
\noalign{\smallskip}
$0048-097$	& A+B & 1.36 & 2.4x1.1 & 46 & 0.08 & 511\\
		& B & 4.88 & 2.2x1.9 & -60 & 0.15 & 982 \\
$0118-272$	& A & 1.36 & 3.0x1.2 & -21 & 0.10 & 742\\
		& B & 4.88 & 5.3x1.4 & 36 & 0.15 & 663\\
$0138-097$	& A+B & 1.36 & 2.3x1.5 & 42 & 0.07 & 541\\
		& B & 4.88 & 4.3x1.7 & -53 & 0.15 & 695\\ 
$0426-380$	& A & 1.36 & 4.4x1.0 & 23 & 0.10 & 624 \\
		& B & 4.88 & 5.3x1.4 & 15 & 0.15 & 1371 \\
$0454+844\ast$	& W & 1.40 & 9.2x26.8 & 57 & 0.20 & 1~~(310)\\
$0537-441$	& A & 1.36 & 5.3x1.2 & 1 & 0.50 & 3010 \\
		& B & 4.88 & 7.1x1.3 & -3 & 1.80 & 5138 \\  
$0716+714\ast$	& W & 1.40 & 17.5x11.0 & 0 & 0.30 & 83~~(280)\\
$0814+425\ast$	& W & 1.40 & 19.5x13.5 & 22 & 0.25 & 10~~(1099) \\
$0820+225\ast$	& W & 1.40 & 33.7x13.8 & 12 & 0.30 & 81~~(1988)\\
$0823+033\ast$	& D & 1.36 & 54.0x43.0 & -52 & 0.15 & 1479\\
$0828+493$	& B$^{\bf a}$ & 1.39 & 5.4x5.4 & 0 & 0.13 & 355 \\
~~~~~~~~~~~~~~$\ast$& W & 1.39 & 18.3x13.1 & 52 & 0.3 & 10~~(288)\\
$0851+202\ast$	& W & 1.40 & 39.4x12.5 & 3 & 0.30 & 2~~(1143) \\
$0954+658$	& A & 1.36 & 1.1x1.0 & 34 & 0.08 & 597 \\
		& B & 4.88 & 2.8x1.3 & 49 & 0.15 & 523 \\
$1144-379$	& A & 1.36 & 3.6x1.1 & -9 & 0.5 & 21~~(1892) \\
$1147+245$	& A+B & 1.36 & 1.3x1.3 & 0 & 0.07 & 796\\
		& B & 4.88 & 2.6x1.9 & 82 & 0.10 & 845 \\
$1308+326$	& B & 1.46 & 4.3x4.3 & 0 & 0.15 & 859 \\
$1418+546$	& B$^{\bf a}$ & 1.40 & 5.4x5.4 & 0 & 0.15 & 566\\
		& W & 1.40 & 17.0x12.0 & 14 & 0.25 & 19~~(710) \\
		& D & 1.36 & 68.0x32.0 & -69 & 0.07 & 815 \\
$1514-241$	& A+B & 1.36 & 3.0x2.0 & 28 & 0.15 & 1606 \\
		& B & 4.88 & 3.0x1.5 & -50 & 0.17 & 2918\\
		& D$^{\bf b}$& 1.40 & 45x45 & 0 & 0.40 & 1993\\
$1519-273\ast$	& A & 1.36 & 3.4x1.4 & -30 & 0.2 & 1690 \\
~~~~~~~~~~~~~~$\ast$& B & 1.36 & 9.3x3.2 & 33 & 0.10 & 883\\
$1652+398$	& B & 1.36 & 3.5x3.0 & -84 & 0.13 & 1383\\
		& B & 4.88 & 1.6x1.3 & -76 & 0.18 & 1320 \\
$1749+701\ast$	& W & 1.40 & 18.3x11.6 & -61 & 0.20 & 1~~(650) \\
$1803+784$	& B & 1.36 & 9.3x4.9 & -11 & 0.07 & 1761 \\
		& B & 4.88 & 2.5x1.2 & -13 & 0.30 & 2214 \\
$1807+698$	& A & 1.36 & 3.2x1.5 & -49 & 0.15 & 1136 \\
		& A+B+D & 1.36 & 7.0x5.5 & -22 & 0.10 & 1284 \\
		& B & 4.88 & 2.5x1.3 & -9 & 0.15 & 1507 \\
$1823+568\ast$	& W & 1.40 & 30.0x10.0 & -53 & 0.30 & 50~~(1190) \\
$2007+777$	& W & 1.40 & 24.0x10.0 & -47 & 0.25 & 17~~(1095) \\
$2131-021$	& A & 1.36 & 1.6x1.5 & 1 & 0.20 & 1291 \\
		& B & 4.88 & 1.9x1.4 & -22 & 0.20 & 1669\\
$2240-260$ 	& A & 1.36 & 2.0x1.0 & 6 & 0.08 & 813 \\
		& B & 4.88 & 3.7x1.3 & -29 & 0.20 & 805\\
$2254+074\ast$	& D & 1.36 & 47.0x42.0 & 4 & 0.15 & 284 \\
\noalign{\smallskip}
\hline
\noalign{\smallskip}
\end{tabular}
\end{center}
\end{table*}

\section{Results}

The VLA and WSRT observations detected extended emission around about 85\% 
of the observed objects; for about 50\%  of them  the flux density of the extended emission we measured is a factor 1.3-3 
higher than previously derived from images with lower dynamic range
(Antonucci \& Ulvestad 1985, Antonucci et al. 1986, Murphy et al. 1993, Kollgaard et al. 1992, Kollgaard et al. 1996).
The WSRT and VLA D-array data were effective on the detection of arc-minute 
scale structures, but only in a few cases we observed significant emission at 
this scale above a surface brightness of $\sim$1 mJy/beam.

The images are shown in Fig. 1 through 34 and the image parameters are reported 
in Table 2. More information on this table is given in the caption. For all the images the contour levels are -3, 3, 6, 12, 25, 50, 
100, 200, 400,  800, 1500, 3000 times the image r.m.s..  

The relevant result for our study is the radio luminosity of the 
extended structure, but we have derived other important physical 
parameters from the images.
The flux density of the extended emission has been determined by subtracting 
the peak from the total flux density, using the most suitable 
image, i.e. the image with enough resolution to isolate the core,
but still containing all the extended emission.
To calculate the luminosity we used the luminosity distance obtained from 
the redshift. For sources having only a lower limit to the redshift we 
used this lower value.

Table 3 contains the flux density and luminosity of the extended emission,
the largest angular and linear size, the core prominence parameter R and the ``sidedness'' of the extended emission of the source: 1 if one-sided, 2 if two-sided, P for a pointlike source.
The luminosities were K-corrected by the factor (1+z)$^{\alpha_{\rm ext}-1}$, 
where $\alpha_{\rm ext}$ is the spectral index of the 
extended structure (S $\propto \nu^{-\alpha}$) and we have used $\alpha_{\rm ext}=0.8$. In Table 3 we also report the flux densities of the extended emission,
as found in the literature, of the remaining 6 sources from the 1 Jy sample 
and of a few sources also observed by us,  whenever the measure
found in the literature had a better resolution allowing a more
accurate subtraction of the core emission. When
necessary, the measures have been scaled to 1.36 GHz.

The largest angular and linear sizes have been calculated only for the sources
showing a well defined extended structure. For the one-sided 
objects we give the largest angular distance from the core. 

The core prominence parameter, R, is defined as  the ratio between the core and
the extended emission flux densities, multiplied by
(1+z)$^{-\alpha_{\rm ext}}$,
to apply the K-correction, assuming $\alpha_{\rm core}=0$ for the core
spectral index. 

In the WSRT images, the cross marks the position
of the subtracted core component.
For 0454+844, 0814+425, 0820+225, 0851+202, 1749+701, 1823+568, 
and 2254+074, observed with the WSRT, and 0823+033 (VLA, D configuration) we 
could not evaluate the contribution of the core and the extended components, 
due to the relatively low resolution of these arrays; for 0716+714 our measure of the extended emission flux density is lower than that found by Antonucci et al. (1986) due
to the low resolution of our data. No image is shown for these sources.

We give now a brief description for each source that showed extended 
structure and a comparison with previous observations.\\
~\\
{\bf 0048-097}: Figure 1  shows the A+B configuration image. 
The core emission is located at the center of a diffuse structure and a bright 
knot (hot-spot?) is seen at the southern edge of the radio source. 
The 4.88 GHz image (Fig. 2) from the B array does not show all the extended 
structure seen at lower frequency. The image at 1.46 GHz published by Wardle 
et al. (1984) only shows the core and a hint of the southern knot. We find a 
flux density of 176 mJy in the extended emission to be compared to the 95 mJy 
reported by Antonucci \& Ulvestad (1985) based on the image of Wardle et al. 
(1984). 0048-097 appears only marginally resolved in our D array data.\\
~\\
{\bf 0118-272}:  
the radio image from the A array (1.36 GHz in Fig. 3) shows a halo around the 
core that contributes significantly to the total luminosity, while the B 
configuration image (not shown) displays only a marginally resolved structure. 
In the image at 4.88 GHz (Fig. 4) there is a hint of a jet extending towards 
SE. Previous observations of this source are reported by Perley (1982).
Our higher dynamic range allows to image the extended emission to larger
distances than in Perley (1982); the diffuse emission in our image extends to
18 arcseconds.\\
~\\
{\bf 0138-097}: 
in the combined A and B array data sets at 1.36 GHz (Fig. 5) we see a weak 
elongated structure leading to the NW and a rather bright component 
extending approximately 4'' south of the core. The 4.88 GHz image (Fig. 6) does 
not show clear secondary components, maybe just a clue of the southern one. 
The source is point-like in the D array image. Previous observations (Perley, 
1982) do not reveal any extended structure in this source. 

We remark that recent optical observations by Heidt et al. (1996) and Scarpa
et al. (1999) show a number of companions within 3 arcseconds from the
BL Lac. The extended radio emission in Fig. 5 is aligned with the
position of these companions.
We cannot rule out that at least part of the extended radio emission
we report in Table 3 is associated with any of these structures.\\
~\\
{\bf 0426-380}: 
this source has a short jet pointing to the NW visible only in
the higher resolution images (Fig. 7 and 8). 
We find a flux density of the extended emission twice the value reported by 
Perley (1982). \\
~\\ 
{\bf 0537-441}: 
the low declination of the source resulted in a very elongated beam in N-S. 
Nevertheless in our highest resolution image (Fig. 9, A array, 1.36 GHz) 
0537-441 shows a curved jet-like structure leading to the west. 
The radio source is slightly resolved also in the B array image at the same
wavelength (not shown). 
The jet has been also detected at 4.88 GHz, but its surface brightness is 
lower (Fig. 10). Perley (1982) detects only the knot at the end of the jet. \\
~\\
{\bf 0828+493}:
no extended emission has been detected for this source either in our WSRT 
image  or in previous VLA observations (Murphy et al. 1993).
However, some extended emission has been recently revealed by the FIRST survey. We used this image (Fig. 11) to derive the numbers in Table 3.\\
~\\
{\bf 0954+658}: 
Kollgaard et al. (1992) detect a jet extending approximately 5'' to the South.
Our A array image at 1.36 GHz (Fig. 12) and the 4.88 GHz B array image 
(Fig. 13) show an elongated structure directed to the SW and then bending to 
the south. No further extended emission has been detected in our B and D 
array data at 1.36 GHz.\\
~\\
{\bf 1144-379}:   
the radio image in the L band is dominated by an unresolved component of 1.9 Jy. 
Perley (1982) reports an upper limit of 10 mJy for any extended emission.
The sources 1144-379 is barely resolved even in the VLA A
array images. 
A coarse estimate of the extended flux, through fitting the radio source with
a point-like component and determining the residuals, gives a value of
about 20 mJy. 
In Fig. 14 we show the VLA image in which the arcsecond core
has not been restored (its position is marked by a cross). Given the low
declination of the source, the uv-coverage is not adequate to allow a proper
imaging of this additional component.\\
~\\
{\bf 1147+245}:
we detect a diffuse  component up to 15'' to the South and another diffuse 
component located about 10'' on the opposite 
side of the core (Fig. 15), basically in agreement with the image published 
by Antonucci \& Ulvestad (1985). Our measure of the extended emission flux 
density is about twice their value. In the 4.88 GHz image (Fig. 16) only 
hints of the extended structures are visible. \\
~\\
{\bf 1308+326}:
The VLA data of this source are from a different program, in which 1308+326
was observed as secondary calibrator.
This source shows a dominant component which is resolved in the SE direction, 
a secondary component $\sim$12'' north, and a diffuse halo surrounding the 
entire structure (Fig. 17). Murphy et al. (1993) resolve the southern diffuse 
emission in a structure suggesting a helical jet. Our B array image reveals 
$\sim 30\%$ more flux density in the extended structure than reported by 
Murphy et al. (1993). \\
~\\
{\bf 1418+546}:
Murphy et al. (1993) found a component to the West of the compact core also 
present in the image from the FIRST survey (Fig. 18). Our WSRT observations 
(Fig. 19) show that this component is elongated to the south, 
with an extended flux density higher than 
reported by Murphy et al. or revealed by the FIRST image. The low 
resolution D array 
image (Fig. 20) is dominated by a point-like component; however extended
and diffuse emission is detected all around, suggesting the presence of
a halo with a total size of 4.5 arcminutes similar to that observed in
1807+698 (see below).\\
~\\
{\bf 1514-241}:
Antonucci \& Ulvestad  (1985) report a component 21'' away from the core. 
Our A+B array image at 1.36 GHz shows a jet emerging
along the SE direction and bending towards NE after a dozen of arcseconds, for
a total extent $\sim55''$ (Fig. 21), also visible in the 4.88 GHz
image (Fig. 22). The image from the D array (Fig. 23) was obtained from the 
NVSS survey (Condon et al. 1998), and clearly shows a diffuse
emission on the arcmin scale on the same side of the jet. 
The flux density of this component was added to the estimate from our A+B 
array image.\\
~\\
{\bf 1519-273}:
this is a very compact source, unresolved in our images at the resolution of 
1''. The source is still unresolved at the mas scale (O'Dea et al. 1991, 
Shen et al. 1997). Perley (1982) gives an upper limit of about 5 mJy for any
extended emission. 
We did
not detect significant extended 
emission above 0.5 mJy/beam, and we do not show any image for this
source.\\
~\\
{\bf 1652+398}: Kollgaard et al. (1992) find 
a diffuse emission at 5 GHz which is in agreement with the 75'' wide halo we
detect in our higher resolution 
B array image at 1.36 GHz (Fig. 24). Our 4.88 image shows instead only
a small fraction of the extended emission visible in the L band.\\
~\\
{\bf 1803+784}:
Kollgaard et al. (1992) find a weak component 45'' away from the core, while 
Antonucci et al. (1986) detect a diffuse emission around the core. 
In our image (Fig. 26) a jet-like structure is present connecting the 
secondary component to the core, and additional diffuse emission on the west 
side.
The 4.88 GHz image (not reported) shows only an unresolved core.\\
~\\
{\bf 1807+698}:
Kollgaard et al. (1996) find the radio structure of this source, at arcsecond 
resolution, consisting of a $\sim$30'' long jet extending from
the core to the west direction. Wrobel \& Lind (1990) find a double lobed structure of total extension of $\sim 60''$, at 4.88 GHz (VLA, B configuration). 
Our image from A+B+D configuration (Fig. 27) shows a diffuse halo of 
$\sim$220'' of extension surrounding the core-jet structure in agreement with the morphology seen by Wrobel \& Lind (1990). 
The jet is clearly visible in our A array image at 1.36 GHz (Fig. 28) and in 
the B array 4.88 GHz image (Fig. 29).\\
~\\
{\bf 2007+777}:
the WSRT image  has not enough resolution to characterize the 
extended structure, which is better highlighted by the VLA observations of 
Murphy et al. (1993). However the total flux density of the extended emission 
in our WSRT image exceeds the measure from Murphy et al. by about 30\% 
(Fig. 30).\\
~\\
{\bf 2131-021}:
the unresolved core is located at the NW edge of the radio emission 
(Fig. 31). Two jet like structures are oriented in P.A. $\sim -$90$^{\circ}$ 
and P.A. $\sim -$170$^{\circ}$; all this is reminiscent of NAT/WAT morphology. 
The B array image at 1.36 GHz does not reveal any further extended emission 
and we can therefore consider that the total angular size of the extended 
radio emission is about 9''. Observations of this source were made by Perley 
(1982)  and he found an extended flux of about 50 mJy. 
Recently a VLA image in B configuration at 1.49 GHz  was published by 
Hutchings et al. (1998). The lower resolution of their observations did not 
allow to properly separate the core flux density from the tail-shaped extended
emission yielding to underestimate the total extended flux density.\\
~\\
{\bf 2240-260}:
in this object the unresolved core sits in the center of a diffuse extended 
emission which can be characterized by two misaligned and bent jet-like 
structures (Fig. 33). The total size of the extended emission is about 26'' 
corresponding to 212 kpc at the redshift of the host galaxy. 
No previous arcsecond scale observations have been found in the literature.\\
~\\
\begin{figure}[b] 
\caption[]{{\bf 0048-097}, VLA A+B configuration, 1.36 GHz. The restoring beam is 2.4x1.1 arcsec in PA 46\deg. The peak flux density is 511 mJy/beam and the r.m.s. noise on the image is 0.08 mJy/beam}
\end{figure}
\begin{figure}[p] 
\caption[]{{\bf 0048-097}, VLA B configuration, 4.88 GHz. The restoring beam is 2.2x1.9 arcsec in PA -60\deg. The peak flux density is 982 mJy/beam and the r.m.s. noise on the image is 0.15 mJy/beam }
\end{figure}
\begin{figure}[p] 
\caption[]{{\bf 0118-272}, VLA A configuration, 1.36 GHz. The restoring beam is 3.0x1.2 arcsec in PA -21\deg. The peak flux density is 742 mJy/beam and the r.m.s. noise on the image is 0.10 mJy/beam  }
\end{figure}
\begin{figure}[p] 
\caption[]{0118-272, VLA B configuration, 4.88 GHz. The restoring beam is 5.3x1.4 arcsec in PA 36\deg. The peak flux density is 663 mJy/beam and the r.m.s. noise on the image is 0.15 mJy/beam }
\end{figure}
\begin{figure}[p] 
\caption[]{0138-097, VLA A+B configuration, 1.36 GHz. The restoring beam is 2.3x1.5 arcsec in PA 42\deg. The peak flux density is 541 mJy/beam and the r.m.s. noise on the image is 0.07 mJy/beam}
\end{figure}
\begin{figure}[p] 
\caption[]{0138-097, VLA B configuration, 4.88 GHz. The restoring beam is 4.3x1.7 arcsec in PA -53\deg. The peak flux density is 695 mJy/beam and the r.m.s. noise on the image is 0.15 mJy/beam}
\end{figure}
\begin{figure}[p] 
\caption[]{0426-380, VLA A configuration, 1.36 GHz. The restoring beam is 4.4x1.0 arcsec in PA 23\deg. The peak flux density is 624 mJy/beam and the r.m.s. noise on the image is 0.10 mJy/beam}
\end{figure}
\begin{figure}[p] 
\caption[]{0426-380, VLA B configuration, 4.88 GHz. The restoring beam is 5.3x1.4 arcsec in PA 15\deg. The peak flux density is 1371 mJy/beam and the r.m.s. noise on the image is 0.15 mJy/beam}
\end{figure}
\begin{figure}[p] 
\caption[]{0537-441, VLA A configuration, 1.36 GHz. The restoring beam is 5.3x1.2 arcsec in PA 1\deg. The peak flux density is 3010 mJy/beam and the r.m.s. noise on the image is 0.50 mJy/beam}
\end{figure}
\begin{figure}[p] 
\caption[]{0537-441, VLA B configuration, 4.88 GHz. The restoring beam is 7.1x1.3 arcsec in PA -3\deg. The peak flux density is 5138 mJy/beam and the r.m.s. noise on the image is 1.80 mJy/beam}
\end{figure}
\clearpage
\begin{figure}[p] 
\caption[]{0828+493, VLA B configuration, 1.40 GHz (from FIRST, Becker et al. 1995). The restoring beam is 5.4x5.4 arcsec. The peak flux density is 355 mJy/beam and the r.m.s. noise on the image is 0.13 mJy/beam}
\end{figure}
\begin{figure}[p] 
\caption[]{0954+658, VLA A configuration, 1.36 GHz. The restoring beam is 1.1x1.0 arcsec in PA 34\deg. The peak flux density is 597 mJy/beam and the r.m.s. noise on the image is 0.08 mJy/beam}
\end{figure}
\begin{figure}[p] 
\caption[]{0954+658, VLA B configuration, 4.88 GHz. The restoring beam is 2.8x1.3 arcsec in PA 49\deg. The peak flux density is 523 mJy/beam and the r.m.s. noise on the image is 0.15 mJy/beam}
\end{figure}
\begin{figure}[p] 
\caption[]{1144-379, VLA A configuration, 1.36 GHz. The restoring beam is 3.6x1.1 arcsec in PA -9\deg. The peak flux density is 21 mJy/beam and the r.m.s. noise on the image is 0.5 mJy/beam}
\end{figure}
\begin{figure}[p] 
\caption[]{1147+245, VLA A+B configuration, 1.36 GHz. The restoring beam is 1.3x1.3 arcsec. The peak flux density is 796 mJy/beam and the r.m.s. noise on the image is 0.07 mJy/beam}
\end{figure}
\begin{figure}[p] 
\caption[]{1147+245, VLA B configuration, 4.88 GHz. The restoring beam is 2.6x1.9 arcsec in PA 82\deg. The peak flux density is 845 mJy/beam and the r.m.s. noise on the image is 0.10 mJy/beam}
\end{figure}
\begin{figure}[p] 
\caption[]{1308+326, VLA B configuration, 1.46 GHz. The restoring beam is 4.3x4.3 arcsec. The peak flux density is 859 mJy/beam and the r.m.s. noise on the image is 0.15 mJy/beam}
\end{figure}
\begin{figure}[p] 
\caption[]{1418+546, VLA B configuration, 1.40 GHz (from FIRST, Becker et al. 1995). The restoring beam is 5.4x5.4 arcsec. The peak flux density is 566 mJy/beam and the r.m.s. noise on the image is 0.15 mJy/beam}
\end{figure}
\clearpage
\begin{figure}[p] 
\caption[]{1418+546, WSRT, 1.40 GHz. The restoring beam is 17.0x12.0 arcsec in PA 14\deg. The peak flux density is 19 mJy/beam and the r.m.s. noise on the image is 0.25 mJy/beam}
\end{figure}
\begin{figure}[p] 
\caption[]{1418+546, VLA D configuration, 1.36 GHz. The restoring beam is 68.0x32.0 arcsec in PA -69\deg. The peak flux density is 815 mJy/beam and the r.m.s. noise on the image is 0.07 mJy/beam}
\end{figure}
\begin{figure} 
\caption[]{1514-241, VLA A+B configuration, 1.36 GHz. The restoring beam is 3.0x2.0 arcsec in PA 28\deg. The peak flux density is 1625 mJy/beam and the r.m.s. noise on the image is 0.15 mJy/beam}
\end{figure}
\begin{figure}[p] 
\caption[]{1514-241, VLA B configuration, 4.88 GHz. The restoring beam is 3.0x1.5 arcsec in PA -50\deg. The peak flux density is 2918 mJy/beam and the r.m.s. noise on the image is 0.17 mJy/beam}
\end{figure}
\begin{figure} 
\caption[]{1514-241, VLA D configuration, 1.40 GHz (from NVSS, Condon et al. 1998). The restoring beam is 45.0x45.0 arcsec. The peak flux density is 1993 mJy/beam and the r.m.s. noise on the image is 0.40 mJy/beam}
\end{figure}
\begin{figure}[p] 
\caption[]{1652+398, VLA B configuration, 1.36 GHz. The restoring beam is 3.5x3.0 arcsec in PA -84\deg. The peak flux density is 1383 mJy/beam and the r.m.s. noise on the image is 0.13 mJy/beam}
\end{figure}
\begin{figure}[p] 
\caption[]{1652+398, VLA B configuration, 4.88 GHz. The restoring beam is 1.6x1.3 arcsec in PA -76\deg. The peak flux density is 1320 mJy/beam and the r.m.s. noise on the image is 0.18 mJy/beam}
\end{figure}
\clearpage
\begin{figure}[p] 
\caption[]{1803+784, VLA B configuration, 1.36 GHz. The restoring beam is 9.3x4.9 arcsec in PA -11\deg. The peak flux density is 1761 mJy/beam and the r.m.s. noise on the image is 0.07 mJy/beam}
\end{figure}
\begin{figure*} 
\caption[]{1807+698, VLA A+B+D configuration, 1.36 GHz. The restoring beam is 7.0x5.5 arcsec in PA -22\deg. The peak flux density is 1284 mJy/beam and the r.m.s. noise on the image is 0.10 mJy/beam}
\end{figure*}
\begin{figure} 
\caption[]{1807+698, VLA A configuration, 1.36 GHz. The restoring beam is 3.2x1.5 arcsec in PA -49\deg. The peak flux density is 1136 mJy/beam and the r.m.s. noise on the image is 0.15 mJy/beam}
\end{figure}
\begin{figure} 
\caption[]{1807+698, VLA B configuration, 4.88 GHz. The restoring beam is 2.5x1.3 arcsec in PA -9\deg. The peak flux density is 1507 mJy/beam and the r.m.s. noise on the image is 0.15 mJy/beam}
\end{figure}
\begin{figure}[p] 
\caption[]{2007+777, WSRT, 1.40 GHz. The restoring beam is 24.0x10.0 arcsec in PA -47\deg. The peak flux density is 17 mJy/beam and the r.m.s. noise on the image is 0.25 mJy/beam}
\end{figure}
\begin{figure}[p] 
\caption[]{2131-021, VLA A configuration, 1.36 GHz. The restoring beam is 1.6x1.5 arcsec in PA 1\deg. The peak flux density is 1291 mJy/beam and the r.m.s. noise on the image is 0.20 mJy/beam}
\end{figure}
\begin{figure}[p] 
\caption[]{2131-021, VLA B configuration, 4.88 GHz. The restoring beam is 1.9x1.4 arcsec in PA -22\deg. The peak flux density is 982 mJy/beam and the r.m.s. noise on the image is 0.20 mJy/beam}
\end{figure}
\begin{figure}[p] 
\caption[]{2240-260, VLA A configuration, 1.36 GHz. The restoring beam is 2.0x1.0 arcsec in PA 6\deg. The peak flux density is 813 mJy/beam and the r.m.s. noise on the image is 0.08 mJy/beam}
\end{figure}
\begin{figure}[p] 
\caption[]{2240-260, VLA B configuration, 4.88 GHz. The restoring beam is 3.7x1.3 arcsec in PA -29\deg. The peak flux density is 805 mJy/beam and the r.m.s. noise on the image is 0.20 mJy/beam}
\end{figure}

\begin{table*}
\caption{ Observational data of the whole 1Jy sample. The question marks
in columns [2] through [7] indicate
indicate a very uncertain value. Column [1] IAU name; [2] red shift: Stickel et al. (1993) for all sources but 0138-097 and 0454+844 (Stocke \& Rector 1997), 0814+425 (Falomo et al. 1997) and 2131-021 (Drinkwater et al. 1997); [3] extended flux, where necessary scaled to 1.36 GHz; [4] Log of extended luminosity; [5] maximum diameter of the source in arcsec and [6] in kpc; [7] ratio core/extended flux density; [8] sidedness: `1' one-sided extended emission, `2' double-sided extended emission, `P' unresolved or barely resolved
source; [9] references for the information given in this table: 1 this paper, 2 Murphy et al. (1993),
3 Antonucci et al. (1996), 4 Antonucci \& Ulvestad (1995), 5 Kollgaard et al. (1992),
6 Perlman \& Stocke (1994), 7 FIRST survey.}
\begin{center}
\begin{tabular}{clrrrrrcc}
[1]&~~[2]&[3]~&[4]~~~~&[5]~~~&[6]~~&[7]~~~~&[8]~~&[9]\\
\noalign{\smallskip}
\hline
\hline
\noalign{\smallskip}
Name & ~~~z & S$_{\rm ext}$~ & ~~Log L$_{\rm ext}$ & LAS~ & Size~~~ & R~~~~ & S~~& Ref.\\
& & ~~(mJy) & ~~~(W Hz$^{-1}$) & ~~~~(arcsec) & (kpc)~~ &  & & \\
\noalign{\smallskip}
\hline
\noalign{\smallskip}
0048$-$097 & $>$0.2 	&176~ & $>$25.50~~~ & 24~~~& $>$101~~~& $<$2.5~~~ & 1& 1\\
0118$-$272 & $>$0.557 	&168~ & $>$26.40~~~ & 18~~~& $>$133~~~ & $<$3.0~~~ & 1& 1\\
0138$-$097 & ~~0.733 	&50~ & 26.13~~~ & 23~~~	& 186~~~ & 6.8~~~ & 2& 1\\
0235+164  & ~~0.940 	&36~ & 26.21~~~ & 13~~~ & 110~~~ &  18.2~~~ & 1 &2\\
0426$-$380 & $>$1.030 	&86~ & $>$26.67~~~  & 6~~~ & $>$51~~~ & $<$4.1~~~ & 1& 1\\
0454+844 & $>$1.34	& ...~~& ...~~~ & ...~~~ & ...~~~ & ...~~~ & ... & ...\\
0537$-$441 & ~~0.896 	& 220~ & 26.95~~~ & 13~~~& 109~~~ & 8.2~~~ & 1& 1\\
0716+714 & $>$0.2 	&387~ & $>$25.85~~~ & 18~~~ & $>$76~~~ & $<$0.87~~~ & 1 &3 \\
0735+178 & $>$0.424	&24~ & $>$25.31~~~ & 10~~~ & $>$66~~~ &  $<$70.0~~~ & 1 &2\\
0814+425 & $>$0.6 	&76~ & $>$26.12~~~ & 6~~~ & $>$46~~~ & 16.8~~~ & 1 & 2\\
0820+225 & ~~0.951 	&700~ & 27.51~~~ & 12~~~& 101~~~ &  1.3~~~ & 1 &2\\
0823+033 & ~~0.506? 	&5~ & 24.79?~~ & 18~~~& 129?~~ &191.0~~~& 1 & 2 \\
0828+493 & ~~0.548? 	&25~ & 25.56?~~ & 42~~~& 331?~~ & 10.0~~~ & 2& 7\\
0851+202 & ~~0.306 	&17~ & 24.87~~~ & 28~~~ & 156~~~ &  81.6~~~ & 1 &6\\
0954+658 & ~~0.367 	&34~ & 25.33~~~ & 6~~~	& 37~~~ & 14.7~~~ & 1& 1\\
1144$-$379 & ~~1.048 	&10? & 25.75?~~ & 1?~~ & 8?~~ & 106.5?~~ & P& 1\\
1147+245 & $>$0.2 	&50~ & $>$24.96~~~ & 34~~~ & $>$144~~~ & $<$13.7~~~& 2&1\\
1308+326 & ~~0.997 	&105~ & 26.73~~~ & 31~~~ & 264~~~ &  4.7~~~ & 1 & 1\\
1418+546 & ~~0.152 	&47~ & 24.69~~~ & 75~~~ & 259~~~ & 13.5~~~  & 1& 1\\
1514$-$241 & ~~0.049 	&210~ & 24.34~~~ & 270~~~ & 354~~~ & 7.4~~~ & 1 & 1 \\
1519$-$273 & $>$0.2 	&...~ & ...~~ & ...~~~ & ...~~~& ...~~~ & P & 1\\
1538+149 & ~~0.605	&234~ & 26.62~~~ & 7~~~ & 53~~~ & 4.5~~~& 1 &2 \\
1652+398 & ~~0.033 	&95~ & 23.65~~~ & 75~~~& 68~~~ & 14.2~~~ & 2& 1\\
1749+096& ~~0.320	& ...~ & ...~~ & ...~~~ & ...~~~ & ...~~~ & ... & ...\\
1749+701 & ~~0.770	&12~ & 25.55~~~	& 3~~~ & 24~~~ & 32.2~~~ & ? & 5 \\
1803+784 & ~~0.684 	&43~ & 26.00~~~ & 56~~~	& 444~~~ & 27.0~~~ & 1& 1\\
1807+698 & ~~0.051 	&1010~ & 25.06~~~ & 222~~~& 302~~~& 1.2~~~ & 2 & 1\\
1823+568 & ~~0.664 	&525~ & 27.06~~~ & 25~~~& 196~~~ & 1.3~~~ & ? & 2\\
2005$-$489 & ~~0.071	& ...~ & ...~~ & ...~~~ & ...~~~ & ...~~~ & ... & ...\\
2007+777 & ~~0.342 	&41~ & 25.35~~~ & 28$^a$~~& 166$^a$~~ & 21.1~~~ & 2$^a$&1\\
2131$-$021 & ~~1.285 	&182~ & 27.20~~~ & 9~~~& 77~~~ &  3.7~~~  & 1& 1\\
2200+420  & ~~0.069	&40~  & 23.92~~~ & 15~~~ & 27~~~ & 78.4~~~ & 1 &4\\
2240$-$260 & ~~0.774 	&333~ & 26.99~~~ & 26~~~ & 212~~~ &  1.6~~~ & 2 & 1 \\
2254+074 & ~~0.190 	&17~ & 24.44~~~ & 18~~~ & 73~~~ &  23.2~~~& 1 & 4 \\
\noalign{\smallskip}
\hline
\noalign{\smallskip}
\multicolumn{9}{l}{Note}\\
\multicolumn{9}{l}{a. Value calculated from the image of Murphy et al. (1993)}\\
\end{tabular}
\end{center}
\end{table*}
\clearpage

\section{Discussion and conclusions}
 
Unified Schemes Models consider the low-power FR~I radio galaxies as
the parent, unbeamed population of BL Lac objects. The boundary
between FR~I and the powerful FR~II radio galaxies has been found to
lie around 2$\times 10^{25}$ W Hz$^{-1}$ at 178 MHz (Fanaroff \&
Riley 1974). At higher frequencies, like at 1.36 GHz, the segregation
in radio power is not as evident as at 178 MHz, and in the literature
one commonly finds that $\sim$10$^{24.5}$ W Hz$^{-1}$ is considered the
borderline. Indeed, it is well known (e.g. Bridle 1984; 1987) that
this boundary at 1.4 GHz is not really sharp, although one can infer
that $most$ of the FR~I radio galaxies are below that limit, while
$most$ of the FR~II lie above. 
We therefore investigated the properties of the extended emission in
the 1~Jy BL Lac sample, and in particular we studied the extended,
unbeamed emission.

The monochromatic luminosities derived from our observations are often
larger than those found in the literature; furthermore, for a few
sources we have revealed extended structure previously unknown.
Therefore this work is more complete and goes deeper than previous
efforts in this direction.

In Fig. 35 we present a plot of the distribution of the luminosity of
the extended, unbeamed emission for the 1 Jy BL Lac sample, as given
in Table 3,
once the arcsecond scale core has been subtracted out. 
We have used the measurements derived from our observations, completed
with data taken from the literature (see Table 3). On this respect, we
chose the better suited images/measurements in order to have about the
same accuracy  in the subtraction of the arcsecond core emission.
~\\
\begin{figure}[h] 
\caption[]{Numeric distribution of the radio luminosity of
the extended emission at 1.36 GHz for the 1 Jy sample, as from Table 3.  The
symbol ``$>$'' is a lower limit to the luminosity, as derived from the
minimum redshift. When the radio luminosity could not be determined
with enough accuracy (0823+033, 0828+493 and 1144-379, see text), we
used the symbol ``?'' to remark the uncertain estimate.}  
\end{figure}
~\\

The unbeamed radio luminosities at 1.36 GHz clearly extend beyond the
limit (not really sharp at this frequency) of 10$^{24.5}$ W Hz$^{-1}$
separating FR~I from FR~II radio galaxies, with just a few
objects with radio luminosities typical of the brightest
FR~II galaxies.  In general the luminosity distribution of the whole
sample seems rather smooth and covers a wide range of radio
luminosities. 

In a forthcoming paper we will discuss in detail
our results, and we will compare the radio properties of the Radio
Selected BL Lacs (i.e. the 1 Jy sample) with those of X-ray Selected BL
Lacs (i.e. EMSS sample) and further with those of FR~I and FR~II radio
galaxies.

\section{Summary}

We have presented interferometric radio data at 1.36, 1.66 and 4.8 GHz
with the VLA at A, B and D configurations, and at 1.4 GHz with the
WSRT, on 28 of 34 objects of the 1 Jy  sample of BL Lac objects (Stickel
et al. 1991). We have obtained high sensitivity radio images at the
arcsec resolution, in order to evaluate the  extended luminosity of
these objects, most of which had poor or no radio data at this
resolution. We found that most of the sources observed possess
substantial extended emission, and often our flux densities exceed
those previously reported into the literature.
A few sources have unbeamed radio luminosities at 1.36 GHz of the
order of 10$^{27}$ W Hz$^{-1}$, supporting the hypothesis that the 
parent population of BL Lac objects is a mixture of FR~I and FR~II
radio sources.  

\acknowledgements
{DD and MB acknowledge financial support from the European Union as an EU
Fellow under contract CHBGCT920212. 
This work was partly supported by the Italian Ministry for University 
and Research (MURST) under grant Cofin98-02-32
The National Radio Astronomy Observatory is operated by Associated
University Inc under contract with the National Science Foundation.
The WSRT is operated by the Netherlands Foundation for Research in
Astronomy (NFRA) with financial support by the Netherlands
Organization for Scientific Research (NWO).
We thank the referee, Travis Rector, whose comments significantly
improved the paper.
~\\ 
~\\
~\\

\end{document}